\documentclass[a4paper,11pt]{article}

\usepackage{graphicx,hyperref}
\usepackage{amsmath}
\usepackage{amssymb}
\usepackage{authblk}
\usepackage{fancyhdr}
\usepackage[titletoc]{appendix}
\usepackage{multicol}
\usepackage{subcaption}
\usepackage{bm}
\usepackage{amsfonts}
\usepackage{epstopdf}
\usepackage{xcolor}
\usepackage{color}
\usepackage{wasysym}
\usepackage{hhline}
\hypersetup{
	colorlinks,    citecolor=blue,    filecolor=blue,    linkcolor=blue,    urlcolor=blue
}

\title{\sffamily{Magnetic field amplification in a laser-irradiated thin foil \\by return current electrons carrying orbital angular momentum}}

\author[1]{Y. Shi}
\author[1]{K. Weichman}
\author[2]{R. J. Kingham}
\author[1]{A. V. Arefiev}

\affil[1]{Department of Mechanical and Aerospace Engineering, University of California at San Diego, La Jolla, CA 92093, USA}
\affil[2]{Blackett Laboratory, Imperial College London, London SW7 2AZ, United Kingdom}

\date{\today}% It is always \today, today,

\begin{document}

\vskip -2.0cm
\maketitle
\vskip -3.0cm
%\vskip -2.0cm

\begin{abstract}
Magnetized high energy density physics offers new opportunities for observing magnetic field-related physics for the first time in the laser-plasma context. We focus on one such phenomenon, which is the ability of a laser-irradiated magnetized plasma to amplify a seed magnetic field. We performed a series of fully kinetic 3D simulations of magnetic field amplification by a picosecond-scale relativistic laser pulse of intensity $4.2\times 10^{18}$~W/cm$^2$ incident on a thin foil. We observe axial magnetic field amplification from an initial 0.1~kT seed to 1.5~kT over a volume of several cubic microns, persisting hundreds of femtoseconds longer than the laser pulse duration. The magnetic field amplification is driven by electrons in the return current gaining favorable orbital angular momentum from the seed magnetic field. This mechanism is robust to laser polarization and delivers order-of-magnitude amplification over a range of simulation parameters.
\end{abstract}

%++++++++++++++++++++++++++++++++++++++++++++++++++++++++++++++++++
\section{\sffamily{Introduction}}

High energy density physics (HEDP) emerged as a new sub-field only about two decades ago, but has already substantially advanced our understanding of materials under extreme conditions and led to the development of several applications, such as those involving energetic particle beams. Much of this progress is due to breakthroughs in laser technology which have enabled drivers capable of depositing energy on a picosecond time scale and creating the high energy density state of matter~\cite{Danson2019}. Until recently, quasi-static magnetic fields have been of relatively low importance in HEDP research due to the technological challenges associated with generating sufficiently strong macroscopic fields at the laser facilities used for HEDP research. 

The recent development of open-geometry, all-optical magnetic field generators~\cite{fujioka2013coil,santos2015coil,Law2016apl} which are portable to any high-energy laser facility has opened up new regimes relevant to magnetized HEDP to exploration. It is now feasible to experimentally probe laser-plasma interactions with an embedded magnetic field that reaches hundreds of Tesla in strength. As a result, there has been an increased interest in laser-driven, high-energy-density systems embedded in strong magnetic fields. Such systems may deliver advances in inertial confinement fusion~\cite{Chang2011prl, Strozzi2012pop, Perkins2013pop}, particle sources~\cite{Macchi2013rmp, Arefiev2016}, and atomic physics~\cite{Dong2001rmp}. 

Concurrently, limitations on the strength of externally applied magnetic fields have also stimulated research into mechanisms of magnetic field generation and amplification by the laser-irradiated plasma itself. One such well-known method for generating quasi-static magnetic axial fields in an underdense plasma is the inverse Faraday (IF) effect enabled by circularly polarized light~\cite{Sheng1996, Haines2001, Najmudin2001, Ali2010}. An underdense plasma with net orbital angular momentum (OAM) can also be created using intense twisted laser beams rather than circularly polarized lasers~\cite{Shi2018,Vieira2018,Nuter2018}. Magnetic field generation has also been observed in solid density targets, including the generation of azimuthal surface and bulk magnetic fields due to the propagation of relativistic electron beams \cite{Davies1999,Robinson2014, huang2019}, and is beneficial for hot electron transport and electron beam collimation~\cite{Sheng1996, Najmudin2001, Gorbunov1996, Sheng1998, Gorbunov1997, Kaymak2016}. Magnetic field amplification has also been observed in laser-produced plasma, for example in shocks \cite{meinecke2014turbulent}, colliding flows \cite{tzeferacos2018laboratory}, implosions \cite{Wessel1986, Chang2011,Nakamura2018}, and with twisted light \cite{Wu2017}. 

In this paper, we introduce a novel method for the amplification of a seed axial magnetic field in the interaction of a picosecond-scale, relativistic intensity laser pulse with a thin solid foil. This mechanism persists for both linear and circular polarization and can amplify a seed axial magnetic field of 20-100~T by a factor of 10.
This paper is organized as follows: In Section~\ref{sec:polns}, we demonstrate magnetic field amplification for three choices of laser polarization, which distinguishes our amplification mechanism from the inverse Faraday effect. In Section~\ref{sec:jtheta}, we show that the magnetic field amplification is associated with the azimulthal current we observe in simulations, which in Section~\ref{sec:OAM} we identify as originating from the favorable orbital angular momentum electrons gain in the return current. In Section~\ref{sec:scan}, we explore the robustness of the amplification mechanism to the choice of simulation parameters.

%++++++++++++++++++++++++++++++++++++++++++++++++++++++++++++++++

\section{\sffamily{Observation of axial magnetic field amplification}} \label{sec:polns}

\begin{figure*}
\centering
\includegraphics[width=0.9\columnwidth]{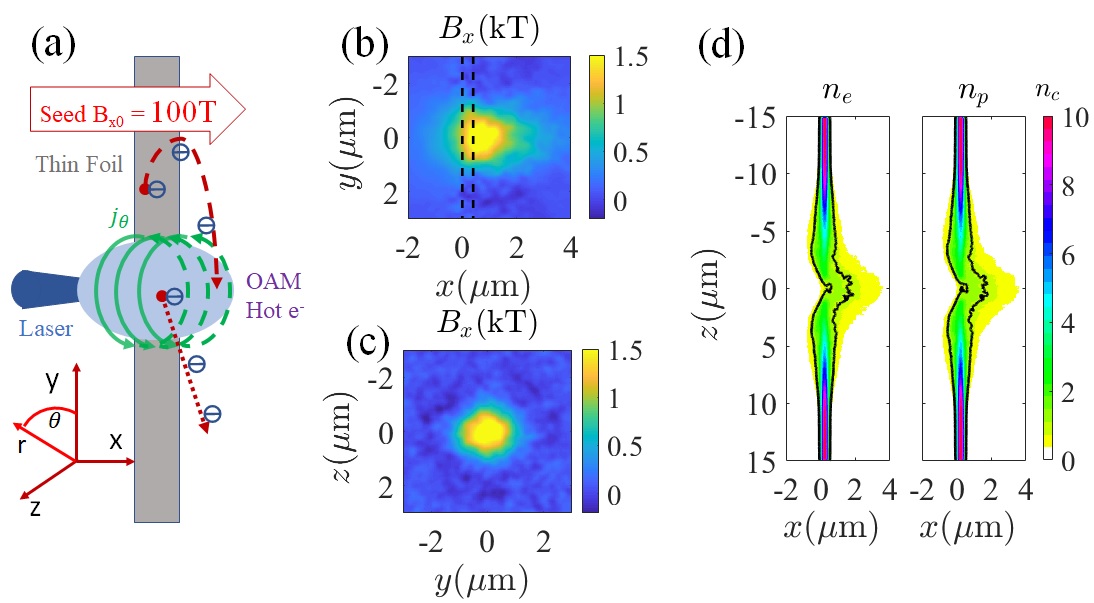}
\caption{Magnetic field amplification in a thin laser-irradiated foil. (a)~Schematic of axial magnetic field amplification. Initially, the laser pondromotively expels electrons creating charge separation. Later, electrons in the return current gain favorable OAM in the seed magnetic field, creating an azimuthal current which amplifies the field. (b) and (c)~Amplified axial magnetic field at $t=340$~fs in the (b)~$x$-$y$ plane ($z$ = 0 $\mu$m, dashed lines denote the original foil position), and (c)~$y$-$z$ plane ($x$ = 0.5 $\mu$m). $B_{x}$ is temporally (16 fs) and spatially averaged in the plane with stencil size 0.25~$\mu$m$\times$0.25~$\mu$m. (d)~Electron (left) and proton (right) density at $t=340$~fs. The black contours denote the critical density $n_{c}$.} \label{SchemeB}
\end{figure*} 

We conduct 3D simulations of a picosecond-scale, relativistic intensity laser pulse interacting with a thin foil of fully ionized hydrogen with an imposed uniform axial seed magnetic field $B_{x0}$. We take the laser pulse to be either circularly or linearly polarized while keeping the same peak intensity.
Simulations were carried out using the fully relativistic particle-in-cell code EPOCH~\cite{arber2015contemporary}. Detailed parameters for the simulation are given in Table \ref{table:PIC}.

\begin{table}
\centering
\begin{tabular}{ |l|l| }
  \hline
  \multicolumn{2}{|l|}{\textbf{Laser parameters} }\\
  \hline
  Peak intensity & $4.2 \times 10^{18}$ W/cm$^2$ \\
  Normalized field amplitude & \\
  Circular polarization & $a_{y0}=a_{z0} = 1.0$ \\
  Linear polarization & $a_{y0}= 1.414$ \\
  Wavelength & $\lambda = 0.8$ $\mu$m \\
  Pulse duration ($\sin^2$ electric field) & $\tau_g=600$ fs\\
  Focal spot size ($1/\mathrm{e}$ electric field) & 3 $\mu$m \\
  Location of the focal plane & $x = -2$ $\mu$m\\
  Laser propagation direction & $+x$ \\
  \hline \hline
  \multicolumn{2}{|l|}{\textbf{Other parameters} }\\
  \hline
  Foil thickness & 0.4 $\mu$m\\
  Electron density & $n_e = 10\;n_{c}$ \\  
  Seed magnetic field strength & $B_{x0}=0.1$ kT \\
  Transverse size of simulation box & $30$ $\mu$m $\times30$ $\mu$m\\  
  Spatial resolution & 40 cells/$\mu$m \\
  Macroparticles per cell for each species &  2 \\
  \hline \hline
  \multicolumn{2}{|l|}{\textbf{Position and time reference}} \\
  \hline
  Location of the front of the foil & $x = 0$ \\
  Time when peak of the laser is at $x=0$ & $t = 0$ \\
  \hline
   \end{tabular}
  \caption{3D PIC simulation parameters. $n_{c} = 1.8\times 10^{21}$~cm$^{-3}$ is the critical density corresponding to the laser wavelength.}
  \label{table:PIC}
\end{table}

We observe magnetic field amplification for three different laser polarization configurations, right hand circularly polarized, left hand circularly polarized, and linearly $y$-polarized. The relative strengths of the amplified magnetic field are given in Table~\ref{table:pol-max}. In the right hand circularly polarized case, the seed magnetic field is amplified from the initial $B_{x0} = 0.1$~kT to a peak amplitude of $B_{x} = 1.5$~kT over a volume of 4~$\mu$m$^3$ in approximately 300~fs. Fig.~\ref{SchemeB}(b) and (c) show the axial magnetic field in the $x$-$z$ and $y$-$z$ planes. The dashed lines in Fig.~\ref{SchemeB}(b) correspond to the initial position of the foil. 

\begin{table}
\centering
\begin{tabular}{ |l|l|l| }
  \hline
  \textbf{Polarization} & \textbf{Peak magnetic field} & \textbf{Averaged magnetic field}\\
  \hline
  Right hand circular & 1.5~kT & 0.9~kT \\
  Left hand circular & 0.6~kT & 0.38~kT\\
  Linear & 0.8~kT & 0.45~kT\\
  \hline \hline
  \multicolumn{3}{|l|}{\textbf{No seed magnetic field} $\mathbf{(B_{x0}=0)}$} \\
  \hline
  Right hand circular & 0.4~kT & 0.2~kT\\
  \hline
   \end{tabular}
  \caption{Amplified magnetic field strength for three laser polarizations with initial seed magnetic field $B_{x0}=0.1$~kT and for right hand circular polarization with no seed magnetic field ($B_{x0}=0$).}
  \label{table:pol-max}
\end{table}

By probing different laser polarizations, we find that the observed magnetic field amplification clearly goes beyond the inverse Faraday (IF) effect.
In the IF effect, the spin angular momentum of a circularly polarized laser \cite{Haines2001} or the orbital angular momentum of a twisted laser \cite{Ali2010} is transferred to electrons in the plasma, driving the generation of a magnetic field. In the absence of a seed magnetic field, the right hand circularly polarized laser pulse generates a peak magnetic field of  $B_x = 0.4$~kT. In all cases with a seed magnetic field, we see peak magnetic fields significantly above this level. We additionally observe substantial magnetic field amplification with linear polarization, albeit with broken symmetry in the $y$-$z$ plane (see Appendix~\ref{append:LpDriven}), and with a left hand circularly polarized laser pulse, where the IF effect generates a magnetic field in the opposite direction from the initial seed. 
In light of these simulations, we find it likely that the high magnetic field amplitude we observe in the right hand circularly polarized simulation represents a combination of magnetic field generation via the IF effect and magnetic field amplification via the novel mechanism we describe in this work. 
The remainder of this work will elucidate the magnetic field amplification process incorporating analysis of the right hand circularly polarized case.

\section{\sffamily{Relationship between magnetic field amplification and $j_\theta$}} \label{sec:jtheta}

\begin{figure*}
    \centering
    \begin{subfigure}{0.45\textwidth} 
    \includegraphics[width=0.95\linewidth]{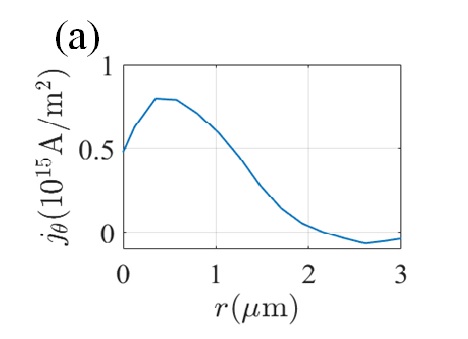}
    \subcaption{}
    \end{subfigure}
    \begin{subfigure}{0.45\textwidth} 
    \includegraphics[width=0.95\linewidth]{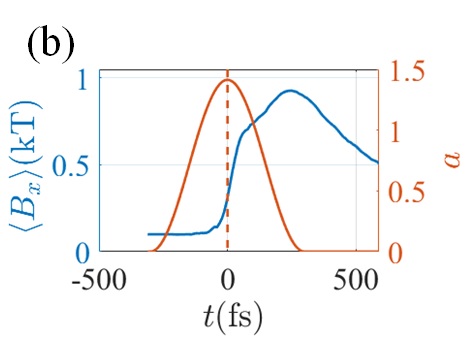}
    \subcaption{}
    \end{subfigure}
    \vspace{-1.75\baselineskip}
\caption{Azimuthal current density, axial magnetic field strength, and reference laser amplitude. (a)~Radial dependence of azimuthal current density at $t=340$~fs, averaged over 0.4~$\mu$m~$<x<0.9$~$\mu$m. (b)~Time evolution of axial magnetic field strength (blue line, left axis) averaged over the cylinder $r< 1$~$\mu$m, $0<x<1$~$\mu$m, with reference laser amplitude $a$ at the foil surface (red line, right axis). The red dashed line denotes $t=0$, which is when the peak of the laser pulse would pass the front foil surface in the absence of the plasma.} \label{JBa}
\end{figure*}

We find that magnetic field amplification is driven by electrons in the return current that arises after the peak of the laser pulse hits the foil surface. When the laser pulse interacts with the foil, it expels electrons from the laser spot, creating charge separation which later induces a return current of electrons relaxing back towards the laser interaction volume to neutralise the space charge. This return current obtains orbital angular momentum (OAM) from the seed axial magnetic field (Fig.~\ref{SchemeB}(a)). The corresponding azimuthal current $j_\theta$ (Fig.~\ref{JBa}(a)) drives the amplification of the seed magnetic field. 

In our simulations, magnetic field amplification is driven by the generation of an azimuthal current by electrons which gain favorable OAM. First, in this Section, we show that the azimuthal current can explain the observed magnetic field amplification. Then, in Section~\ref{sec:OAM}, we demonstrate how this azimuthal current can be generated by the electron return current and can persist over hundreds of femtoseconds.

We find that the azimuthal current $j_\theta$ is responsible for the magnetic field amplification, which we confirm by calculating the axial magnetic field strength from the $j_\theta$ we observe in simulations. We see that a positive azimuthal current density $j_{\theta} \sim 10^{15}$~A/m$^2$ develops at small radius (e.g. Fig.~\ref{JBa}(a)) during the amplification process. We estimate the maximum axial magnetic field generated from $j_\theta$ using the Biot-Savart law \cite{jackson1975electrodynamics}, which we simplify by assuming the current density is uniform over the radial extent $R$ and the longitudinal extent 2$\Delta x$,
\begin{equation}
    B_x^{\mathrm{max}} = \dfrac{\mu_0}{2} \int_0^R \int_{-\Delta x}^{\Delta x} j_{\theta} \dfrac{r^2}{(r^2 + x^2)^{3/2}} \mathrm{d}x \mathrm{d}r = \mu_0 j_{\theta} \Delta x \operatorname{arsinh}(R/\Delta x).
\end{equation}
We estimate $R=\Delta x \sim 1$~$\mu$m, which predicts a maximum magnetic field strength of $B_x^{\mathrm{max}} \sim 1.1$~kT, close to the peak magnetic field strength $\langle B_{x} \rangle$ shown in Fig.~\ref{JBa}(b), which was averaged over a cylindrical volume with radius and length of 1~$\mu$m.

We can also use the azimuthal current density to compare the OAM density obtained in this simulation to that which is produced by twisted lasers. The OAM density of electrons as a function of position $r$ can be written as $l_x = rm_en_ev_{\theta}$, where $m_e$ is the electron mass, $n_e$ is the number density, and $v_{\theta}$ is the effective azimuthal velocity. The effective azimuthal velocity is related to the current density by $v_{\theta} = -j_{\theta}/(|e|n_e)$, which allows us to calculate the OAM density as $l_x = -rm_ej_{\theta}/|e|$. Using $j_\theta$ as given above and the electron density from simulations, $n_e \approx 10^{21}$~cm$^{-3}$, we find that the OAM density is $l_x = - 0.02$~kg/m-s at $r=1$~$\mu$m and $v_\theta \approx 0.02c$. Thus we produce a rotating plasma with electron density 2 orders of magnitude higher and rotating velocity 1 order of magnitude higher than is produced by twisted lasers \cite{Shi2018}. In terms of the energy content, we find that the energy in the magnetic field ($\varepsilon_B = \int B^2/(2 \mu_0) dV \approx 1$~$\mu$J) remains small compared to the kinetic energy of electrons around the amplifying area ($\approx 50$~$\mu$J).

%++++++++++++++++++++++++++++++++++++++++++++++++++++++++++++++++

\section{\sffamily{Relationship between $j_\theta$ and the OAM of the return current}} \label{sec:OAM}

We now illustrate how the azimuthal current is produced by electrons which gain favorable OAM. When the laser pulse interacts with the foil, the ponderomotive force expels electrons from the laser spot. Initially, this creates a net charge separation within the laser spot (see Appendix~\ref{append:Denst_r}), which later induces a return current. This radially inward return current gains favorable OAM from the axial magnetic field and consequently provides the $j_\theta$ responsible for magnetic field amplification.

We use the trajectories of test electrons participating in the magnetic field amplification to illustrate the amplification process. During the amplification, we randomly select electrons at the sampling time $t_s$ from the area in which the magnetic field is amplified ($0.5$~$\mu$m~$<x<1.0$~$\mu$m, $|y|<3$~$\mu$m, $|z|<3$~$\mu$m). We then track the full time history of these electrons to capture where they originated and demonstrate how they can obtain OAM.

The averaged trajectory of 344 electrons selected at time $t_s = 290$~fs is given in Fig.~\ref{eAvgTraj_t}. By analyzing the individual (e.g. the movie in Appendix~\ref{append:Movie}) and averaged transverse position of these electrons ($r_{e}(t)$, the blue line in Fig.~\ref{eAvgTraj_t}(a)), we see that the majority of these electrons originate from radius $r_{e} > 10$~$\mu$m, well outside the laser spot, and move inward to small radius during the falling edge of the laser pulse ($t>0$). These observations identify these electrons as belonging to the return current. 

\begin{figure*}
    \centering
    \begin{subfigure}{0.45\textwidth} 
    \includegraphics[width=0.95\linewidth]{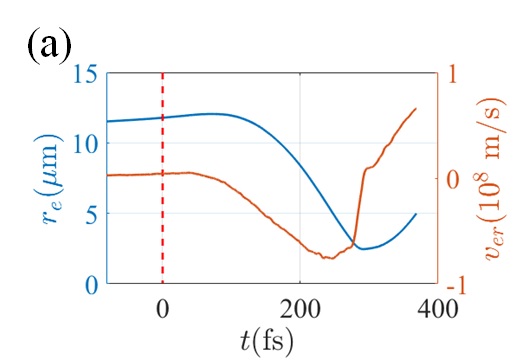}
    \subcaption{}
    \end{subfigure}
    \begin{subfigure}{0.45\textwidth} 
    \includegraphics[width=0.95\linewidth]{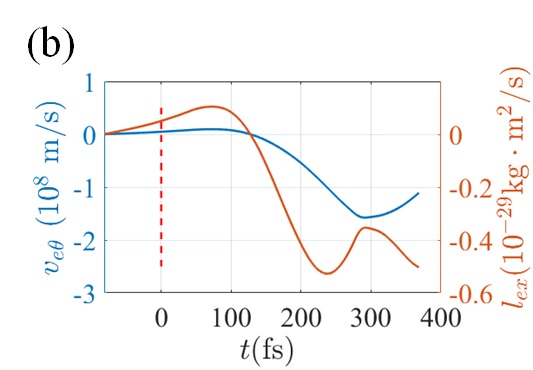}
    \subcaption{}
    \end{subfigure}
    \vspace{-1.75\baselineskip}
\caption{Average trajectory of 344 electrons participating in the amplification process. (a)~Averaged radial position of electrons $r_{e}$ (blue line, left axis) and radial velocity $v_{er}=\mathrm{d}r_{e}/\mathrm{d}t$ (red line, right axis). (b)~Azimuthal velocity which would be expected based on interaction with the seed magnetic field $v_{e\theta} = \int |e|v_{er}B_{x0}/m_e \mathrm{d}t $ (blue line, left axis) and corresponding OAM per electron $l_{ex} = m_ev_{e\theta}r$ (red line, right axis). Red dashed lines indicate $t=0$, the time when the laser pulse is at its peak at $x=0$. Electrons were randomly chosen within the cubic volume $0$~$<x<1$~$\mu$m, $|y|<3$~$\mu$m, $|z|<3$~$\mu$m at the sampling time $t_s=290$~fs.} \label{eAvgTraj_t}
\end{figure*} 

We further see that inward motion of electrons can generate a negative average OAM ($l_{ex} < 0$), which drives magnetic field amplification. The average radial velocity, $v_{er} \equiv \mathrm{d}r_{e}/\mathrm{d}t$, of the test electrons is shown in the red line in Fig.~\ref{eAvgTraj_t}(a). In order to illustrate how the sign of the OAM can be favorable for magnetic field amplification, we consider the azimuthal velocity this radial velocity produces in conjunction with the seed field ($v_{e\theta} = \int |e|v_{er} B_{x0}/m_e dt$), and its corresponding OAM ($l_{ex}=m_e v_{e\theta} r_{e}$). Fig.~\ref{eAvgTraj_t}(b) shows that the negative inward radial velocity associated with the return current can drive $v_{e\theta}<0$ and $l_{ex}<0$, which represents a net positive azimuthal current ($j_{e\theta} >0$) for this group of electrons. The net rotation of electrons in the $y$-$z$ plane is also immediately visible in the trajectories of the electrons, e.g. the movie in Appendix~\ref{append:Movie}.

\begin{figure*}
\centering
\includegraphics[width=0.5\columnwidth]{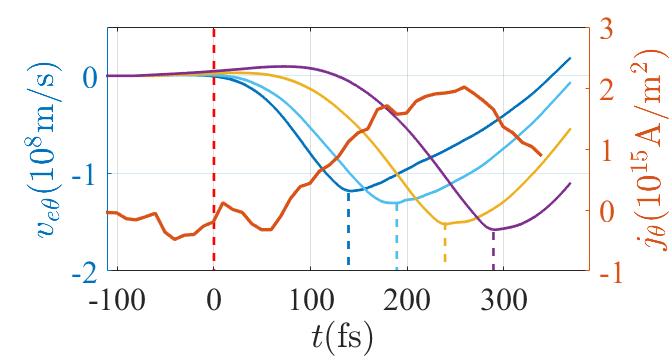}
\caption{Azimuthal velocity for different sets of sampled electrons and azimuthal current density in the plasma. (a)~Azimuthal velocity $v_{e\theta}$ (left axis) calculated as described in Fig.~\ref{eAvgTraj_t} for different sampling times ($t_{s1}=140$~fs, $t_{s2}=190$~fs, $t_{s3}=240$~fs, $t_{s4}=290$~fs), and azimuthal current density (red line, right axis) at location $x=0.4$~$\mu$m and averaged over the circle defined by $r$ = 0.5 $\mu$m. The red dashed line denotes the time when the laser intensity at the front target surface ($x=0$) is maximum.} \label{eTraj_oamt}
\end{figure*}

We replicate the above analysis for groups of electrons chosen at different sampling times, ($t_{s1} = 140$~fs, $t_{s2} = 190$~fs, $t_{s3} = 240$~fs, $t_{s4} = 290$~fs) and find a negative azimuthal velocity ($v_{e\theta}<0$, Fig.~\ref{eTraj_oamt}) can be maintained in the region of magnetic field amplification over a long time, consistent with our observation that a positive azimuthal current density at $r=0.5$~$\mu$m is maintained in the simulation over hundreds of femtoseconds (red line in Fig.~\ref{eTraj_oamt}). Looking earlier in time, we see that the azimuthal current density is negative around the time when the peak of the laser pulse impacts the foil target ($t=0$, red dashed line in Fig.~\ref{eTraj_oamt}). This is consistent with our observation that it is the return current that drives magnetic field amplification.

The net cycle-averaged inward force on an electron starting near rest (i.e. neglecting the magnetic force) encodes the competition between the ponderomotive force of the laser pushing the electron outward and the force due to charge separation pulling the electron inward. During the rising edge of the laser pulse ($t<0$), electrons are pushed predominantly outward, and we see a corresponding small negative azimuthal current density. However, this does not have a substantial impact on the averaged axial magnetic field ($\langle B_x \rangle$ to the left of the red dashed line in Fig.~\ref{JBa}(b)). During the subsequent falling edge of the laser pulse (to the right of the red dashed line in Fig.~\ref{JBa}(b)), the force associated with charge separation can overcome the ponderomotive force of the laser and a net return current is produced. 

We find that the return current is able to drive magnetic field amplification during the falling edge of the laser pulse from $\langle B_{x} \rangle \approx B_{x0} = 0.1$~kT to  $\langle B_{x} \rangle = 0.9$~kT. After $t=240$~fs, by which time the laser pulse has mostly reflected from the foil (see Fig.~\ref{JBa}(b)), the magnetic field begins to slowly decay.

%+++++++++++++++++++++++++++++++++++++++++++++++++++++++++++++++++++++++

\section{\sffamily{Parameter scan}} \label{sec:scan}

\begin{figure*}
    \centering
    \begin{subfigure}{0.45\textwidth} 
    \includegraphics[width=0.95\linewidth]{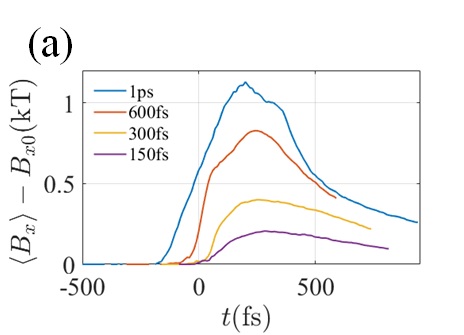}
    \subcaption{}
    \end{subfigure}
    \begin{subfigure}{0.45\textwidth} 
    \includegraphics[width=0.95\linewidth]{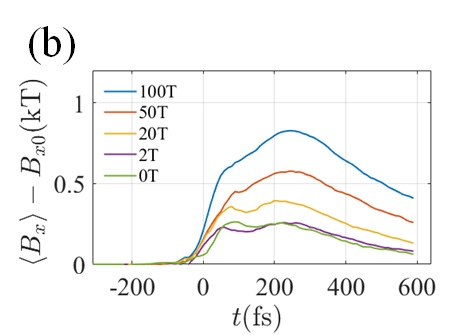}
    \subcaption{}
    \end{subfigure}
    \vspace{-1.75\baselineskip}
\caption{Net amplification of the axial magnetic field. (a)~Scan over laser pulse duration. (b)~Scan over the seed magnetic field strength $B_{x0}$. Only the net amplification, $\langle B_{x} \rangle - B_{x0}$, is shown.} \label{ParaScanTauBx0}
\end{figure*}

We now investigate the robustness of the amplification mechanism described in Sections~\ref{sec:jtheta} and~\ref{sec:OAM} to the choice of simulation parameters. First, we scan the seed magnetic field strength $B_{x0}$ holding all the other parameters the same as in Table \ref{table:PIC} and investigate the balance of magnetic field amplification via our mechanism versus magnetic field generation via the IF effect. 
For a seed magnetic field strength below 20~T, the axial magnetic field becomes indistinguishable from the field produced in the absence of any seed magnetic field (Fig.~\ref{ParaScanTauBx0}(b)). In other words, below 20~T no net magnetic field amplification is seen and the observed magnetic field can be attributed to magnetic field generation via the IF effect.
In contrast, for a seed magnetic field strength above 20~T, we see that the seed magnetic field can be amplified by a factor of 10.

Second, we scan over the laser pulse duration $\tau_g$ at the original seed magnetic field strength ($B_{x0}=0.1$~kT) to probe the impact of pulse duration on the magnetic field amplification. We see that amplification becomes weaker as the driving laser pulse duration becomes shorter (Fig.~\ref{ParaScanTauBx0}(a)). 
We can obtain an average magnetic field as high as 1.1~kT for a pulse duration of 1~ps.

In conjunction with these first two parameter scans, we note that the period of electron Larmor precession in $B_{x0}= 0.1$~kT is around $\tau_B = 340$~fs and that for $B_{x0}= 20$~T, we have $\tau_B = 1700$~fs. This suggests that the amplification process requires the laser pulse duration to be sufficiently long that $\tau_g \gtrsim 0.5\;\tau_B$. 

Third, we consider the robustness of the magnetic field amplification to the foil thickness. The use of a thin foil (0.4~$\mu$m thickness) in our original simulations maintains the feasibility of having the seed axial magnetic field penetrate the target on a reasonable time scale. However, for thicker foils the experimental time scales for generating the seed field and allowing it to penetrate into the foil must be accounted for. For example, experiments have shown that the capacitor coil target can produce an axial magnetic field in excess of $0.1$~kT with a sub-nanosecond rise and slow decay in excess of 10 nanoseconds \cite{fujioka2013coil,santos2015coil,santos2018coil}, which corresponds to approximately a 10~$\mu$m penetration depth into a copper-like conductive material.
Without performing fully self-consistent simulations of the magnetic field penetration into the foil, we are therefore limited to studying $\mu$m-scale foil thickness.
For the sake of demonstrating that magnetic field amplification is feasible in a thicker foil, we consider a second case with 2~$\mu$m thickness. 
We see that magnetic field amplification is still present, albeit at a reduced level.

%++++++++++++++++++++++++++++++++++++++++++++++++++++++++++++++++

\section{\sffamily{Summary}} \label{sec:summary}

We demonstrate a novel mechanism for magnetic field amplification by a short pulse laser interacting with a thin foil capable of amplifying a 0.1~kT seed to 1.5~kT over a spatial extent of several cubic microns and persisting for hundreds of femtoseconds longer than the laser pulse duration. We find that magnetic field amplification is driven by the return current arising during the falling edge of the laser pulse. Electrons in the return current gain orbital angular momentum in the presence of the seed magnetic field, driving an azimuthal current with favorable sign for magnetic field amplification. This amplification process is robust to the choice of simulation parameters and occurs for both linear and circular polarization. For a right hand circularly polarized pulse, we find that a seed magnetic field above 20 T delivers order-of-magnitude amplification from a 600~fs pulse and increasing the pulse duration from 150~fs to 1~ps increases the amplified magnetic field by a factor of 5.

%+++++++++++++++++++++++++++++++++++++++++++++++++++++++

\section{\sffamily{Acknowledgements}}

The work was supported by the DOE Office of Science under Grant No. DE-SC0018312. Y.S. acknowledges the support of Newton International Fellows Alumni follow-on funding. Simulations were performed using the EPOCH code (developed under UK EPSRC Grants No. EP/G054940/1, No. EP/G055165/1, and No. EP/ G056803/1) using HPC resources provided by the TACC at the University of Texas and the ARCHER UK National Supercomputing Service. The authors would like to thank Dr. J. Santos for stimulating discussions.

%+++++++++++++++++++++++++++++++++++++++++++++++++++++++
%+++++++++++++++++++++++++++++++++++++++++++++++++++++++
%+++++++++++++++++++++++++++++++++++++++++++++++++++++++

\appendix
\section{\sffamily{Magnetic field amplification driven by a linearly polarized laser}} \label{append:LpDriven}

\begin{figure*}
    \centering
    \begin{subfigure}{0.45\textwidth} 
    \includegraphics[width=0.95\linewidth]{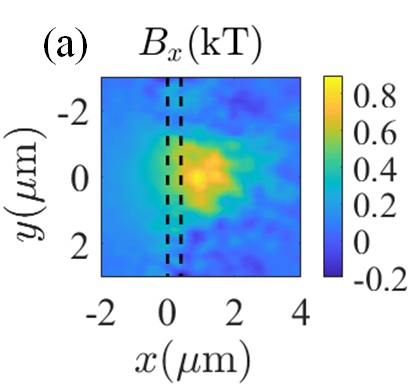}
    \subcaption{}
    \end{subfigure}
    \begin{subfigure}{0.45\textwidth} 
    \includegraphics[width=0.95\linewidth]{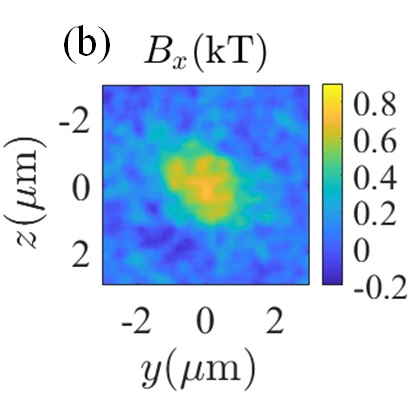}
    \subcaption{}
    \end{subfigure}
    \vspace{-1.75\baselineskip}
\caption{Magnetic field amplification driven by a linear polarized laser at $t=340$~fs. (a)~Axial magnetic field distribution in the $x$-$y$ plane ($z=0$, dashed lines denote the original foil position). (b)~Axial magnetic field distribution in the $y$-$z$ plane ($x=0.5$~$\mu$m).} \label{Lp_Bx2d}
\end{figure*}

In this Appendix, we show magnetic field amplification driven by a linearly $y$-polarized laser pulse. We take $a_{y0} = 1.414$ to obtain the same peak intensity as the circularly polarized case. All other simulation parameters are the same as Table~\ref{table:PIC}. Fig.~\ref{Lp_Bx2d} shows the amplified magnetic field we obtain at $t=340$~fs. Compared to Fig.~\ref{SchemeB}(b) and (c), the magnetic field is weaker in the linearly polarized case. We also see that there is asymmetry in the $y$-$z$ plane (Fig.~\ref{Lp_Bx2d}(b)), which may be related to the polarization direction. 

%+++++++++++++++++++++++++++++++++++++++++++++++++++++++

\section{\sffamily{Charge density, electron current density, and ion current density}} \label{append:Denst_r}

\begin{figure*}
\centering
\includegraphics[width=0.9\columnwidth]{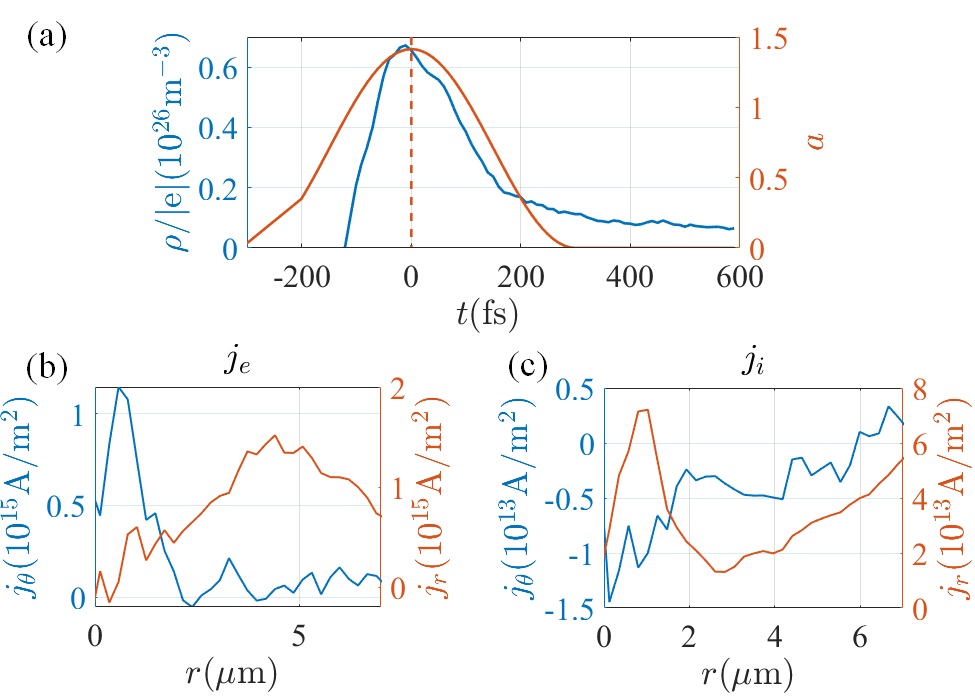}
\caption{Charge density and per-species current density. (a)~Charge density $\rho(t)/|e|$  (blue line, left axis) averaged over $0.2$~$\mu$m~$<x<0.4$~$\mu$m and $1$~$\mu$m~$<r<4$~$\mu$m as a function of time with reference laser amplitude $a(t)$ at the foil surface (red line, right axis). The red dashed line corresponds to $t=0$, when $a(t)$ is maximum. (b)~and (c)~Azimuthal (blue, left axes), and radial (red, right axes) current densities for (b)~electrons and (c)~protons. The current densities are averaged over $0.2$~$\mu$m~$<x<0.4$~$\mu$m.} 
\label{epcDenst}
\end{figure*}

In this Appendix, we present additional properties of the charge density and current present in the thin foil in the right hand circularly polarized case with the simulation parameters given in Table~\ref{table:PIC}. The blue line in Fig.~\ref{epcDenst}(a) shows the charge density distribution $\rho(t)/|e|$ averaged over $0.2$~$\mu$m~$<x<0.4$~$\mu$m and $1$~$\mu$m~$<r<4$~$\mu$m. During the rising edge of the laser pulse, the charge density increases, indicating a net ponderomotive expulsion of electrons from the laser spot.
After the peak of the laser pulse (perpendicular red dashed line in Fig.~\ref{epcDenst}), the ponderomotive force on electrons in the foil, $f_p(t) \propto - \partial a^2 (t) / r\partial r$ decreases, and the charge density $\rho(t)$ also decreased, consistent with a net inward return current.

Fig.~\ref{epcDenst}(b) shows the azimuthal and radial electron current densities (blue and red lines, respectively) as a function of radius at time $t=340$~fs, which is after the laser pulse has been fully reflected by the foil. Fig.~\ref{epcDenst}(c) similarly shows the ion current densities. These densities have been averaged over $0.2$~$\mu$m~$<x<0.4$~$\mu$m. We see that the ion current density is much smaller than the electron current density, which suggests the ion motion in the transverse ($y$-$z$) plane can be ignored. We also note that the total current density is positive, i.e. there is a net inward electron return current.

%+++++++++++++++++++++++++++++++++++++++++++++++++++++++

\section{\sffamily{Movie: trajectories of traced electrons}} \label{append:Movie}

\begin{figure*}
\centering
\includegraphics[width=0.7\columnwidth]{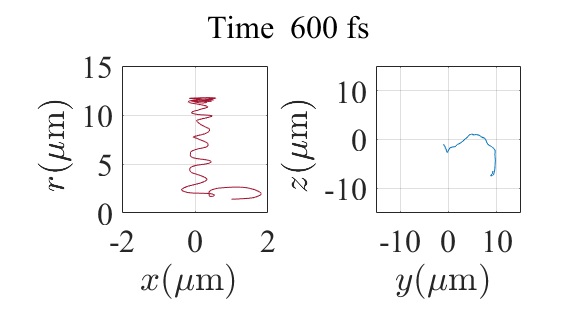}
  \caption{Trajectory of a representative electron. The electron oscillates longitudinally (in $x$) through the foil while moving radially inward (left), and at the same time rotates in the transverse ($y$-$z$) plane (right).} \label{traj1}
\end{figure*}

Fig.~\ref{traj1} shows the trajectory of one of the electrons we traced for the right hand circularly polarized case with the simulation parameters given in Table~\ref{table:PIC}. Electrons oscillate longitudinally through the foil ($x$-direction) while being pulled radially inward (left plot). The inward motion also corresponds to rotation in the transverse ($y$-$z$) plane (right plot). The movie further shows the trajectories of many traced electrons.

%+++++++++++++++++++++++++++++++++++++++++++++++++++++++++

\bibliographystyle{ieeetr}
% \bibliography{aa}

\end{document}